\documentclass[twocolumn,amsmath,amssymb,prb,longbibliography,superscriptaddress,floatfix]{revtex4-1}

\usepackage{graphicx}
\usepackage{calrsfs}
\usepackage{xcolor}
\usepackage{hyperref}

\usepackage[utf8]{inputenc}
\usepackage[T1]{fontenc}
\usepackage{mathptmx}

\begin{document}

\title{Dynamics and Pinning for Skyrmions in Altermagnets}

\author{J. C. Bellizotti Souza}
\email[Corresponding author: ]{jcbsouza@dac.unicamp.br}
\affiliation{``Gleb Wataghin'' Institute of Physics, University of Campinas, 13083-859 Campinas, S\~ao Paulo, Brazil}

\author{C. J. O. Reichhardt}
\author{A. Saxena}
\author{C. Reichhardt}
\affiliation{Theoretical Division and Center for Nonlinear Studies, Los Alamos National Laboratory, Los Alamos, New Mexico 87545, USA}

\date{\today}

\begin{abstract}

We examine the dynamics, Hall angle, and pinning for N{\'e}el skyrmions in an altermagnet. Using an atomistic model, we show that skyrmion velocity and Hall angle dependence is anisotropic with respect to the direction of the drive, due to the fourfold symmetry implied by the two sublattices of the altermagnet. The skyrmion Hall angle and velocity at fixed drive show strong variations for increasing ratios of the exchange constant of the sublattices, $J_2/J_1$. This fourfold anisotropy of altermagnetic (ATM) skyrmions also leads to anisotropic pinning effects for an ATM skyrmion interacting with isotropic circular pinning sites. We also propose a simple particle model for this system that takes into account this anisotropy and find that it captures both the variations of the ATM skyrmion Hall angle and velocity as a function of drive direction, as also found in the atomistic simulations. Using this particle model, we examine ATM skyrmions interacting with a periodic array of pinning sites. For increasing ratios of $J_2/J_1$, we find a strongly non-monotonic ATM skyrmion velocity, where there is a minimum in the velocity where the skyrmion locks to different symmetry directions of the periodic pinning lattice. For a random array, we find that ATM skyrmions show strongly anisotropic depinning thresholds and velocity responses for different drive directions, and that the Hall angle is nearly constant with drive. In comparison, for the same parameters, the depinning threshold for a ferromagnetic (FM) skyrmion is lower, and the skyrmion Hall angle shows a strong velocity dependence. The lower depinning threshold for FM skyrmions is due to stronger Magnus forces.
\end{abstract}

\maketitle

\section{Introduction}

Skyrmions are topological magnetic textures that typically arise in chiral magnets and were first observed
in neutron scattering experiments  \cite{Muhlbauer09} and soon after with direct imaging  \cite{Yu10,Seki12}.
Since then, the field has undergone rapid growth as more systems have been found that can support
skyrmions \cite{Jonietz10,Schulz12,Iwasaki13,Kong13,Nagaosa13,Mochizuki14,Zhang18,EverschorSitte18}.
Skyrmions can be moved with an applied current and via various other methods. They experience a pinning effect upon interacting with quenched disorder \cite{Fernandes18},
so there is typically a current or depinning threshold for skyrmion motion
\cite{Iwasaki13,Lin13,Reichhardt15,Woo16,Juge19}.
Skyrmions can be characterized by a winding number Q. Due to their topology, they can have a strong Magnus or gyroscopic component to their motion \cite{Nagaosa13,EverschorSitte14,Reichhardt15,Jiang17,Legrand17,Litzius17,Zeissler20},
which causes skyrmions to move at an angle with respect to the applied drive, known as the skyrmion Hall angle.
The Magnus effect also influences how skyrmions interact with quenched disorder, which can lead to spiral-like motion of skyrmions around pinning sites.
Due to their size, stability, and mobility, skyrmions are also promising candidates for a number of applications, such as racetrack memory  \cite{Fert13,Tomasello14}
and novel skyrmion-based computing \cite{Pinna20,Song20}.

There are also different types of skyrmions and related magnetic textures that include antiskyrmions
\cite{Hoffmann17,Nayak17}, merons \cite{Yoshimochi24}
antiferromagnetic skyrmions \cite{Barker16,Legrand20,Pham24}, skyrmionium  \cite{Kolesnikov18}, and three-dimensional (3D) Hopfions \cite{Liu20,Zheng23,Souza25a}.
These textures can exhibit distinct dynamics, lattice structures, and
interactions with pinning \cite{Gobel21a,Souza25a}.
For applications, the large skyrmion Hall angle presents a barrier for
creating certain devices. For example, when skyrmions move along a racetrack,
the finite Hall angle causes them to move to the sample edge where they
can annihilate. There have been various efforts to find materials to reduce
the skyrmion Hall effect,
such as by using skyrmionium with a topological charge of $Q = 0$;
however, skyrmionium is unstable at higher drives or strong pinning and breaks up into skyrmions with $Q = 1$ \cite{Kolesnikov18,Souza25}.
One of the most promising systems
in which the skyrmion Hall effect is absent or
small is antiferromagnetic skyrmions
\cite{Barker16,Legrand20,Aldarawsheh24,Yao25},
which can be regarded as two coupled ferromagnetic skyrmions
in which the net topological charge cancels.
Various experiments have been able to realize such textures \cite{Legrand20}
and show that fast motion under a drive with small skyrmion Hall angles
can occur \cite{Dohi19,Pham24}.
In general, there has been growing interest in
antiferromagnets for both basic science and technological applications  \cite{Baltz18}.

Recently, a new type of magnet called an altermagnet has been proposed
\cite{Smejkal22,Smejkal22a,Smejkal23,McClarty24,Jungwirth26},
which has zero macroscopic magnetization but can exhibit
finite higher-order magnetic moments that break Kramers degeneracy. A candidate material is MnTe \cite{Lee24}.
This class of systems has similarities to antiferromagnets, but the symmetry transitions are different, since
altermagnets can be considered as
consisting of two partially overlapping sublattices.
Several experiments have found evidence for altermagnetism  \cite{Krempasky24,Zhang25}.
A natural question is whether altermagnet systems can also host skyrmions and, if so, how such skyrmions would behave under driving and interaction with pinning.
There have already been several investigations of
altermagnet (ATM) skyrmions \cite{Jin24,Vakill25,Schwartz25,Jiang25,Liu26}.
Jin {\it et al.} \cite{Jin24}
found that ATM skyrmions can have a finite skrymion Hall angle due to the formation of an effective quadrupole interaction, and that this Hall angle can depend on the direction of the applied drive.
Jin {\it et al.}~also proposed a Thiele-like equation of motion for ATM skyrmions, which they verified with micromagnetic simulations.
Vakili {\it et al.} \cite{Vakill25}
found that skyrmions in altermagnet systems can exhibit an anisotropic skyrmion Hall effect, while \citet{Schwartz25}
found that ATM skyrmions could be moved with a thermal gradient.
Jiang {\it et al.} \cite{Jiang25}
studied the quadrupolar nature of ATM skyrmions and argued that they could be used as an ideal system to explore systems with $d$-wave symmetry.
Liu {\it et al.} \cite{Liu26}
considered current driving in skyrmions in frustrated altermagnet systems and found that the skyrmion Hall angle does not lock to the skyrmion helicity.
Overall, these studies have provided insights into the unique properties and behaviors of skyrmions in altermagnet systems, which could have important implications for potential applications.

In this work, we consider atomistic simulations of the dynamics and pinning of
%altermagnetic (ATM)
ATM skyrmions. We find that the ATM skyrmion velocity and Hall angle
are strongly dependent on the direction of the driving force, due to the anisotropic nature of the quadrupole moment that arises from the two ATM sublattices
with exchange constants $J_1$ and $J_2$.
For different drive directions, we show how the velocity varies with increasing $J_2/J_1$, where a larger $J_2/J_1$ ratio leads to smaller and
more anisotropic skyrmions.
As $J_2/J_1$ increases, the skyrmion velocity and Hall angle exhibit an anisotropic response.
For ATM skyrmions interacting with isotropic circular pinning sites, we find that the pinning site energy and depinning force are anisotropic.
We also propose a particle-based model of ATM skyrmions, which is simpler than the model of Jin {\it et al.} \cite{Jin24} in that it does not take into account skyrmion distortions, but instead assumes different dissipation components along each axis. This particle model shows strong agreement with the anisotropic velocity and Hall angle response found from the atomistic simulations.
Using this model, we examine the interaction of an ATM skyrmion with a pinning array and find that the pinning efficiency is enhanced at certain values of $J_2/J_1$, where the skyrmion Hall angle is large enough that the skyrmion motion can lock to the different symmetry directions of the pinning lattice.
For ATM skyrmion interactions with random pinning for different drive directions, we find that the depinning threshold and skyrmion sliding velocity are strongly anisotropic, and that the pinning strength is stronger than what is observed
for ferromagnetic skyrmions. We argue that this is due to the lower Magnus force in the ATM system, which causes the gyroscopic motion of the ATM skyrmions
to be too weak to permit the skyrmions to
slide around the pinning sites.
Finally, we discuss some future directions for the study of ATM skyrmion systems.

\section{Simulation}
We consider a two sublattice model,
capable of hosting N{\'e}el ATM skyrmions, governed
by the following Hamiltonian \cite{Jin24}

\begin{align}\label{eq:1}
  \mathcal{H}=&-\sum_{i}J_1\left({\bf m}_{i}^A\cdot{\bf m}_{i+\hat{\bf x}}^A
  +{\bf m}_{i}^B\cdot{\bf m}_{i+\hat{\bf y}}^B\right)\\\nonumber
  &-\sum_{i}J_2\left({\bf m}_{i}^A\cdot{\bf m}_{i+\hat{\bf y}}^A
  +{\bf m}_{i}^B\cdot{\bf m}_{i+\hat{\bf x}}^B\right)\\\nonumber
  &-\sum_{i}J_3{\bf m}_{i}^A\cdot{\bf m}_{i}^B\\\nonumber
  &-\sum_{i, \langle i, j\rangle}\mathbf{D}_{i, j}\cdot\left({\bf m}_{i}^A\times{\bf m}_{j}^A+{\bf m}_{i}^B\times{\bf m}_{j}^B\right)\\\nonumber
  &-\sum_{i}K\left[\left(\hat{\bf z}\cdot{\bf m}_i^A\right)^2 + \left(\hat{\bf z}\cdot{\bf m}_i^B\right)^2\right] \ .
\end{align}

\begin{figure}
  \centering
  \includegraphics[width=\columnwidth]{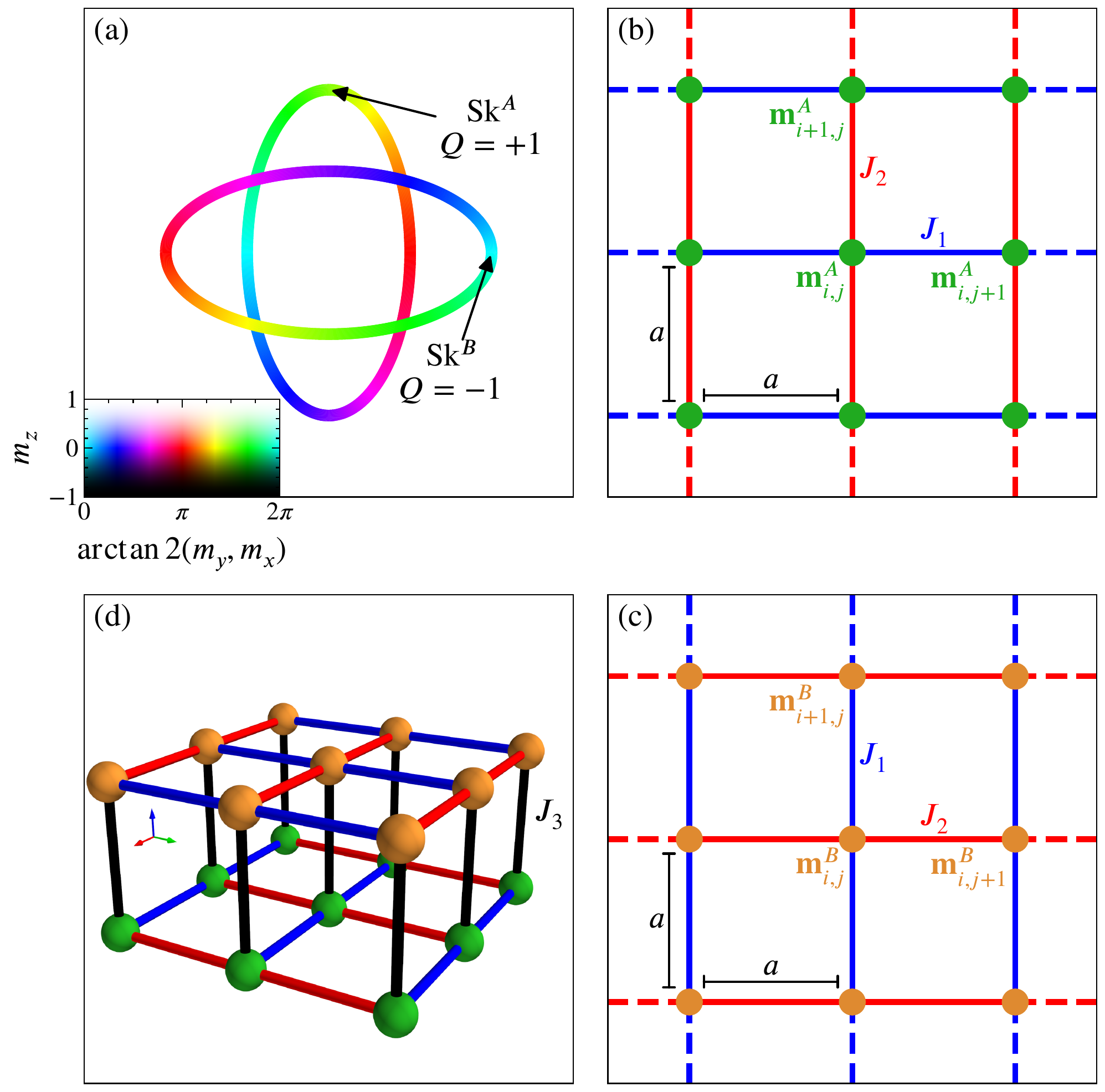}
  \caption{
    (a) Schematic showing ATM skyrmions on sublattices $A$ and $B$.
    Sublattice $A$ ($B$) skyrmions have $Q=+1$ ($Q=-1$),
    so the sample has no net topological
    charge. Inset: colormap used to represent skyrmions throughout
    this work. The colors in the main panel correspond to $m_z=0$.
    As a consequence of the anisotropic exchange interactions,
    the skyrmions
    become elliptical.
    (b) Schematic of sublattice $A$ showing the square
    arrangement of atoms with lattice constant $a$ and the
    magnetic moment
    ${\bf m}_{i, j}^A$ (green). Bonds along $y$ have
    exchange constant $J_2$ (red) and bonds along $x$
    have exchange constant $J_1$ (blue), giving
    an anisotropic exchange interaction.
    (c) The corresponding schematic of sublattice $B$ with
    magnetic moments ${\bf m}_{i, j}^B$ (orange).
    The exchange constants are swapped:
    bonds along $y$ have exchange constant $J_1$ (blue)
    and bonds along $x$ have exchange constant $J_2$ (red).
    (d) Three-dimensional view of a portion of our sample showing
    sublattices $A$ (green) and $B$ (orange) along with exchange
    constants $J_1$ (blue) and $J_2$ (red).
    A third exchange constant $J_3$ (black) couples
    sublattices $A$ and $B$ along $z$.
  }
  \label{fig:1}
\end{figure}

Here $J_1$ is the first exchange constant,
present along $x$ bonds on sublattice $A$
and along $y$ bonds on sublattice $B$. $J_2$
is the second exchange constant, present
along $y$ bonds on sublattice $A$ and
along $x$ bonds on sublattice $B$.
$J_3$ is an antiferromagnetic ($J_3<0$)
coupling between sublattices $A$ and $B$.
Schematics showing the geometry of the exchange interactions
appear in Fig.~\ref{fig:1}(b, c, d).
The next interaction is the interfacial Dzyaloshinskii-Moriya (DM)
interaction, where ${\bf D}_{i, j}=D\hat{\bf z}\times\hat{\bf r}_{i, j}$
is the DM vector and  $\hat{\bf r}_{i, j}$ is the unit
distance vector between sites $i$ and $j$. This
interaction is present only between two sites
of the same sublattice. Here $\langle i, j\rangle$ indicates
that the sum is performed over the first neighbors of the $i$th-site.
The last interaction is the perpendicular magnetic anisotropy (PMA)
with constant $K$.
As shown in Fig.~\ref{fig:1}(b, c, d), the underlying lattice
is a square arrangement of atoms with lattice constant $a$.

The time evolution is obtained by means of the Landau-Lishiftz-Gilbert
equation augmented with external torque contributions \cite{Seki16, Gilbert04, Slonczewski72},

\begin{equation}\label{eq:2}
  \frac{\partial\mathbf{m}^l_i}{\partial
    t}=-\gamma\mathbf{m}^l_i\times\mathbf{H}^\text{eff}_{i, l}
  +\alpha\mathbf{m}^l_i\times\frac{\partial\mathbf{m}^l_i}{\partial t}
  +\boldsymbol{\tau}^l_i\ .
  %  +\frac{pa^3}{2e}\left(\mathbf{j}\cdot\nabla\right)\mathbf{m}_i \ .
\end{equation}
Here $\gamma=1.76\times10^{11}~$T$^{-1}$~s$^{-1}$ is the electron
gyromagnetic ratio, $\mu=\hbar\gamma$ is the atomic magnetic moment,
$l=A, B$ is the sublattice,
$\mathbf{H}^\text{eff}_{i, l}=-\frac{1}{\mu}\frac{\partial \mathcal{H}}{\partial \mathbf{m}^l_i}$
is the effective magnetic field including all interactions from the Hamiltonian, $\alpha$ is the
phenomenological damping introduced by Gilbert, and the last term is
the spin-transfer-torque (STT) contribution caused by application of an in
plane spin polarized current.
We consider interactions between conduction
electrons and our sample in the exchange-dominant
regime, so that STT contributions are given by
\begin{equation}\label{eq:3}
  \boldsymbol{\tau}^l_i=\frac{\hbar\gamma pa^3}{2e\mu}\left[\left(\nabla\cdot{\bf j}\right){\bf m}^l_i-\beta{\bf m}^l_i\times\left(\nabla\cdot{\bf j}\right){\bf m}^l_i\right] \ ,
\end{equation}
where $p$ is the spin polarization, $e$
the electron charge, and $\mathbf{j}=j\left(\cos\phi\hat{\bf x}+\sin\phi\hat{\bf y}\right)$ the applied
current density. The first term is the adiabatic contribution,
and the second term the non-adiabatic contribution, with $\beta$
the non-adiabatic coefficient. We emphasize that $\nabla$ only
acts on a given sublattice $l$;
this also can be seen by ${\bf j}\cdot\hat{\bf z}=0$.
Use of this STT expression implies that the
conduction electron spins are always parallel to the magnetic moments
$\mathbf{m}$ \cite{Iwasaki13,Zang11}, or equivalently,
that we are considering interactions between conduction
electrons and our sample to be in the exchange-dominant
regime \cite{Jin24}.

In this work we fix $J_1=1$~meV, $J_3=-0.05J_1$,
$D=0.2J_1$, $K=0.04J_1$, $a=0.5$~nm, and $\beta=0.005$.
The numerical integration of Eq.~\ref{eq:2} is performed
using a forth order Runge-Kutta method, with an integration
step of $\Delta t=0.01\hbar/J_1\approx0.6$~ps.

%To verify ATM skyrmion deformations, we observe
%${\bf m}$ structure factor, given by
%\begin{equation}
%  \overline{m}_d({\bf q})=\left|\sum_{k, l}\left({\bf m}^l_k\cdot\hat{\bf d}\right)\mathrm{e}^{-{\bf q}\cdot{\bf r_i}}\right| \ ,
%\end{equation}
%where $d$ is the component of ${\bf m}$ begin analyzed,
%and ${\bf q}$ is the wave vector. The sum is performed
%over all atoms of both sublattices.

\section{Results}

\begin{figure}
  \centering
  \includegraphics[width=\columnwidth]{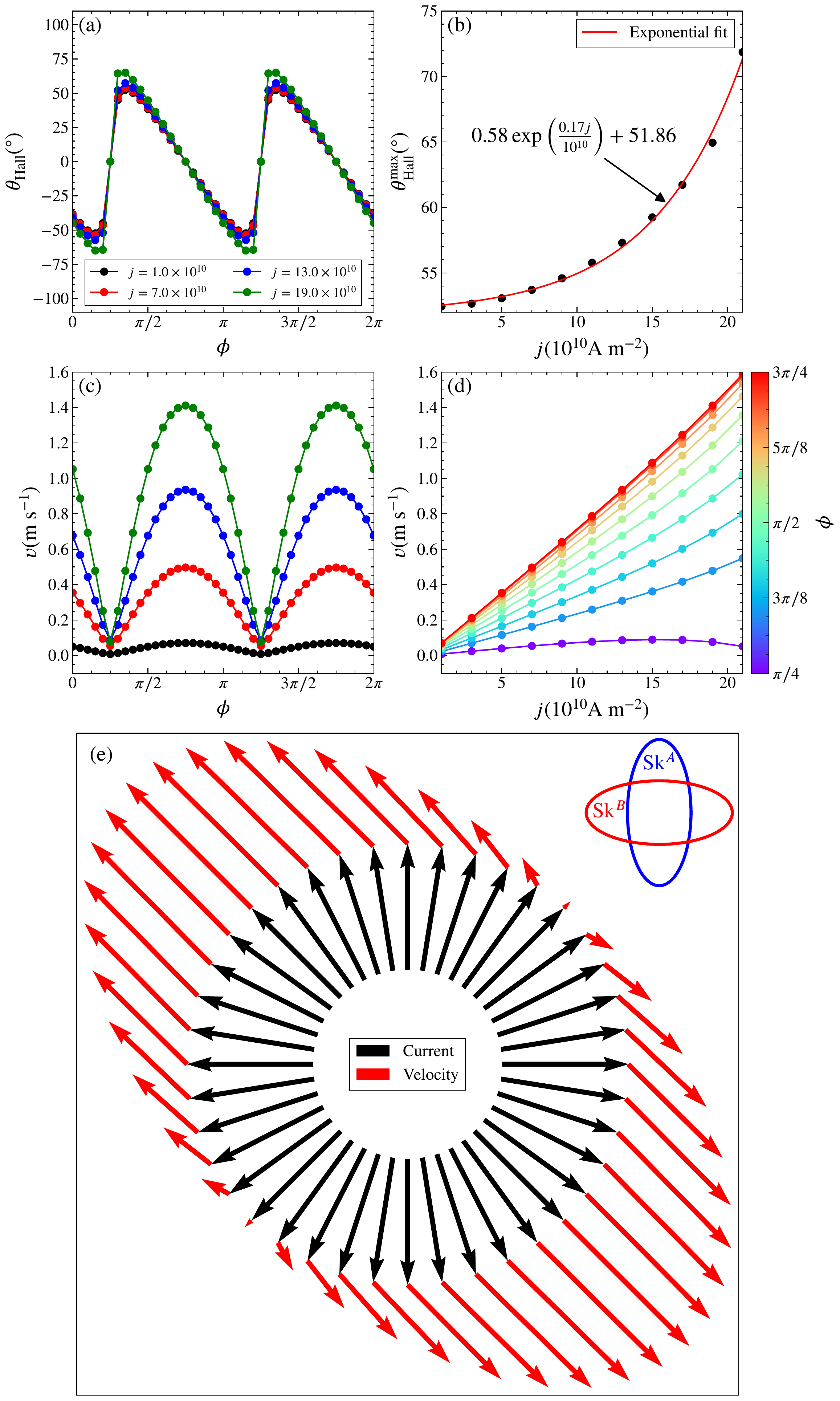}
  \caption{Simulation results for an ATM system with $J_2/J_1=3$.
  (a) The skyrmion Hall angle $\theta_\mathrm{Hall}$ vs current direction $\phi$
  at $j=1.0\times 10^{10}$ (black), $7.0\times 10^{10}$ (red),
  $13.0\times 10^{10}$ (blue), and $19.0\times 10^{10}$ (green).
  (b) The maximum Hall angle $\theta_\mathrm{Hall}^\mathrm{max}$ across
  all $\phi$ values
  vs $j$ (black) along with an exponential fit (red).
 (c) ATM skyrmion velocity, $v=\sqrt{v_x^2+v_y^2}$,
  vs $\phi$ at the same $j$ values from panel (a).
 (d) $v$ vs $j$ at different values of $\phi$.
  (e) Schematic of the current direction
  (black arrows) and the corresponding ATM skyrmion velocity
  (red arrows). Red arrows are in scale
  with respect to one another.
  The inset shows the two skyrmions on each
  sublattice, with nodes occurring where the curves cross.
  }
  \label{fig:2}
\end{figure}

In Fig.~\ref{fig:2}(a) we plot the skyrmion Hall angle $\theta_{\rm Hall}$
versus drive angle $\phi$ for a system with $J_2/J_1=3$ at different
values of the driving current $j$, computed using the two sublattice
model described in Eq.~(1).
Here the ATM skyrmion Hall angle passes through zero at $\phi=\pi/4+n\pi/2$
with $n\in\mathbb{Z}$.
The ATM skyrmion is expected to have 45$^\circ$ symmetry, implying
that currents applied along the directions of the ATM skyrmion nodes
should have an equivalent effect
on the overall texture; however, this is not what we observe.
Instead of the sinusoidal curve found in \citet{Jin24},
Fig.~\ref{fig:2}(a) shows that
there is a sawtooth dependence of $\theta_{\rm Hall}$ on $\phi$.
We extract the
maximum Hall angle value $\theta_\mathrm{Hall}^\mathrm{max}$ from each
of the curves in Fig.~\ref{fig:2}(a)
and plot it as a function of $j$ in Fig.~\ref{fig:2}(b), where we
find that the data are fit
well by an exponential curve.
As for the net velocity
$v=\sqrt{v_x^2+v_y^2}$
of the ATM skyrmion, the plot of $v$ versus $\phi$ in Fig.~\ref{fig:2}(c) at
different currents $j$ shows that $v$ has
an oscillatory behavior, with peak velocities
appearing at $\phi=3\pi/4+n\pi$ and local velocity minima
at $\phi=\pi/4+n\pi$ with $n\in\mathbb{Z}$.
This behavior is the opposite of what is found
for regular ferromagnetic skyrmions, where the skyrmion
velocity is independent of the direction in which the current
is applied.
We plot $v$ versus $j$ for different values of $\phi$ in Fig.~\ref{fig:2}(d).
Here, at $\phi=\pi/4$, $v$ is close to zero.
As $\phi$ increases, the dependence of $v$ on $j$ becomes more linear,
until for the peak velocity values at $\phi=3\pi/4$,
$v$ varies completely linearly with $j$.
To understand this behavior, in Fig.~\ref{fig:2}(e) we plot a schematic of
the dependence of the ATM skyrmion velocity on the direction of the
applied current.
Here we see that ATM skyrmions
only move along $3\pi/4$ or $-\pi/4$,
with the motion angle changing sign at drives aligned with $\phi=\pi/4$
and $\phi=-3\pi/4$.
The inset of Fig.~\ref{fig:2}(e) indicates that
the angles at which the motion angle changes sign match
the directions of the ATM skyrmion nodes.
The peak and minimum velocity values in Fig.~\ref{fig:2}(c) also fall
along the node directions.

\begin{figure}
 \centering
 \includegraphics[width=\columnwidth]{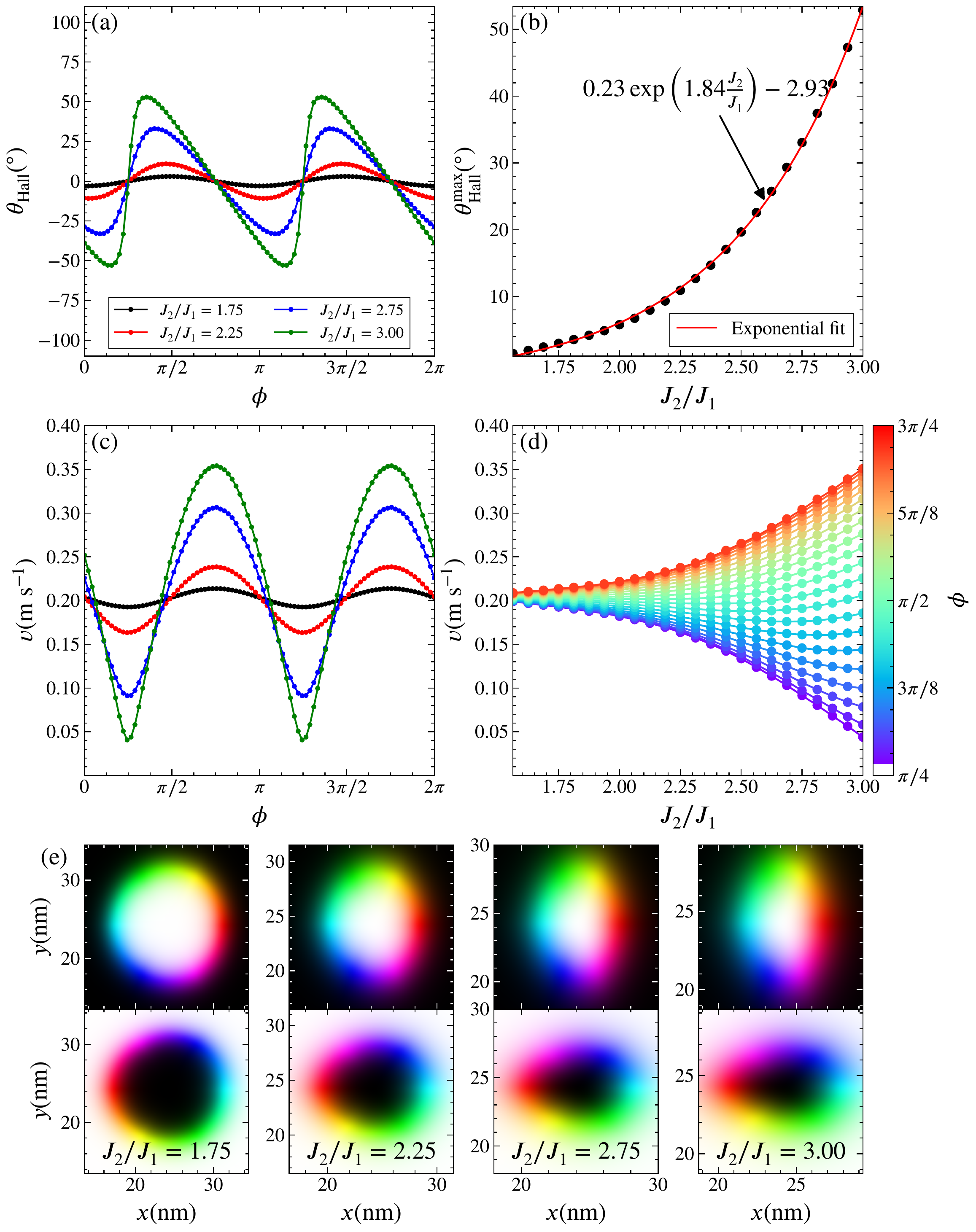}
 \caption{Simulation results for an ATM system with
 $j=5\times10^{10}$~A~m$^{-2}$.
 (a) $\theta_\mathrm{Hall}$ vs $\phi$ at
 $J_2/J_1$=1.75 (black), 2.25 (red), 2.75 (blue), and 3.00 (green).
 (b) The maximum Hall angle $\theta_\mathrm{Hall}^\mathrm{max}$
 across all $\phi$ values vs $J_2/J_1$ (black) along with
 an exponential fit (red).
 (c) Skyrmion velocity $v$ vs $\phi$ at the
 $J_2/J_1$ values from (a).
 (d) $v$ vs $J_2/J_1$ for different $\phi$ values.
 (e) Real space images of different ATM skyrmion configurations
 in the two layers with different $J_2/J_1$.
 }
 \label{fig:3}
\end{figure}

Figure~\ref{fig:3}(a) shows $\theta_{\rm Hall}$ versus $\phi$ at
a fixed current of $j=5 \times 10^{10}$ A m$^{-2}$ for different
$J_1/J_2$ ratios.
We find that $\theta_\mathrm{Hall}$ has
a sawtooth shape
at higher $J_2/J_1$,
but as $J_2/J_1$ decreases, the amplitude of the variations in
$\theta_\mathrm{Hall}$ decreases and the curve becomes more sinusoidal
in shape.
In the $J_2/J_1=1$ limit,
$\theta_\mathrm{Hall}$ approaches a constant
value of $\theta_\mathrm{Hall}=0$.
The skyrmions at
$J_2/J_1=1$ are symmetrical and
have zero net topological charge, corresponding to
an antiferromagnetic skyrmion.
We find that for all $J_2/J_1$ ratios,
$\theta_\mathrm{Hall}$ passes through zero when $\phi=\pi/4+n\pi/2$,
with $n\in\mathbb{Z}$.
The plot of the maximum Hall value
$\theta_\mathrm{Hall}^\mathrm{max}$ versus $J_2/J_1$ in Fig.~\ref{fig:3}(b)
also has an exponential shape.
Comparing Fig.~\ref{fig:3}(b) with Fig.~\ref{fig:2}(b),
we observe a stronger dependence of
$\theta_\mathrm{Hall}^\mathrm{max}$
on $J_2/J_1$ than with $j$, as indicated by the exponential
factors of both fits.
Figure~\ref{fig:3}(c) shows the skyrmion velocity $v$ versus
$\phi$ at different values of $J_2/J_1$.
At higher $J_2/J_1$, $v$ resembles
a rectified sinusoidal curve, while at lower $J_2/J_1$,
$v$ is closer to a pure sinusoidal shape.
The amplitude of the velocity variations as a function of $\phi$
is a strong function of $J_2/J_1$.
In the $J_2/J_1=1$ limit, $v$ approaches a constant value
of $v\approx20$~m~s$^{-1}$. The behavior at lower $J_2/J_1$
is in accordance with the observations in Ref.~\onlinecite{Jin24}.
We find velocity minimums at $\phi=\pi/4+n\pi$,
with $n\in\mathbb{Z}$, and velocity peaks
at $\phi=3\pi/4+n\pi$, with $n\in\mathbb{Z}$.
The minimum velocity values approach $v=0$ as $J_2/J_1$
increases.
We show $v$ versus $J_2/J_1$ for a range of different $\phi$ values
in Fig.~\ref{fig:3}(d).
The peak ATM skyrmion velocity increases
rapidly with
increasing $J_2/J_1$, while the minimum velocity
rapidly decreases towards $v=0$ as $J_2/J_1$ becomes larger.
Real space images of the ATM skyrmion configurations at different
values of $J_2/J_1$ appear in Fig.~\ref{fig:3}(e).
The ATM skyrmions are smaller and more asymmetrical at
higher $J_2/J_1$, and become larger and more symmetrical for
lower $J_2/J_1$.

\citet{Jin24} constructed a quantitatively accurate
model for ATM skyrmions that correctly predicts the behavior of ATM skyrmions,
with results matching those of simulations; however,
below we show that this
is not the simplest model capable of explaining
the asymmetric ATM skyrmion behavior.
Instead, we claim that the simplest model capable of qualitatively reproducing
the asymmetric behavior is one capturing
the combined motion of both skyrmions composing the ATM skyrmion.
We start with the Thiele approximation for rigid skyrmions
\cite{Iwasaki13},
\begin{equation} \label{eq:4}
  G_i\hat{\bf z}\times\left({\bf v}_e-{\bf v}_i\right)+{\bf D}_i\left(\beta{\bf v}_e-\alpha{\bf v}_i\right)+{\bf F}=0 \ ,
\end{equation}
where ${\bf D}_1=\mathcal{D}\begin{pmatrix}1 & \eta \\ \eta & \kappa\end{pmatrix}$
  and ${\bf D}_2=\mathcal{D}\begin{pmatrix}\kappa & \eta \\ \eta & 1\end{pmatrix}$
are the dissipative tensors,
$G$ is the Magnus force strength,
${\bf v}_e$ is the dragging force from STTs,
${\bf v}_i$ is the velocity of each skyrmion component of the ATM skyrmion,
$\alpha$ is the damping,
the $\beta$ term is responsible for
nonadiabatic contributions from STTs,
and ${\bf F}$ is the net force from other interactions.
We are using this model as a toy model, and therefore the values
of these parameters do not directly correspond to those
of our atomistic simulations. For a more
accurate use of this equation, refer to \citet{Iwasaki13}.
To account for the skyrmion asymmetry, we consider ${\bf D}$ to have
different components along each axis represented by
the parameter $\kappa$.
Assuming that
the ATM skyrmion motion is given by ${\bf v}=({\bf v}_1+{\bf v}_2)/2$,
using Eq.~\ref{eq:4} and further assuming $\eta\ll1$, $G_2=-G_1=G$ , and taking
${\bf F}=0$, we obtain
\begin{align}
  \label{eq:5}v_\parallel&=
  \frac{v_e}{2}
  \frac{\left[\mathcal{D}G\left(\beta-\alpha\right)\left(\kappa-1\right)\sin\left(2\phi\right)+2\mathcal{D}^2\alpha\beta\kappa+2G^2\right]}{\mathcal{D}^2\alpha^2\kappa+G^2}\\
  \label{eq:6}v_\perp&=
  \frac{v_e}{2}
  \frac{\mathcal{D}G\left(\beta-\alpha\right)\left(\kappa-1\right)\cos\left(2\phi\right)}{\mathcal{D}^2\alpha^2\kappa+G^2} \ ,
\end{align}
where $v_\parallel$ and $v_\perp$ are the velocities along and
perpendicular to the applied current. We emphasize
that this is a toy model and not a real model of the system.

\begin{figure}
  \centering
  \includegraphics[width=\columnwidth]{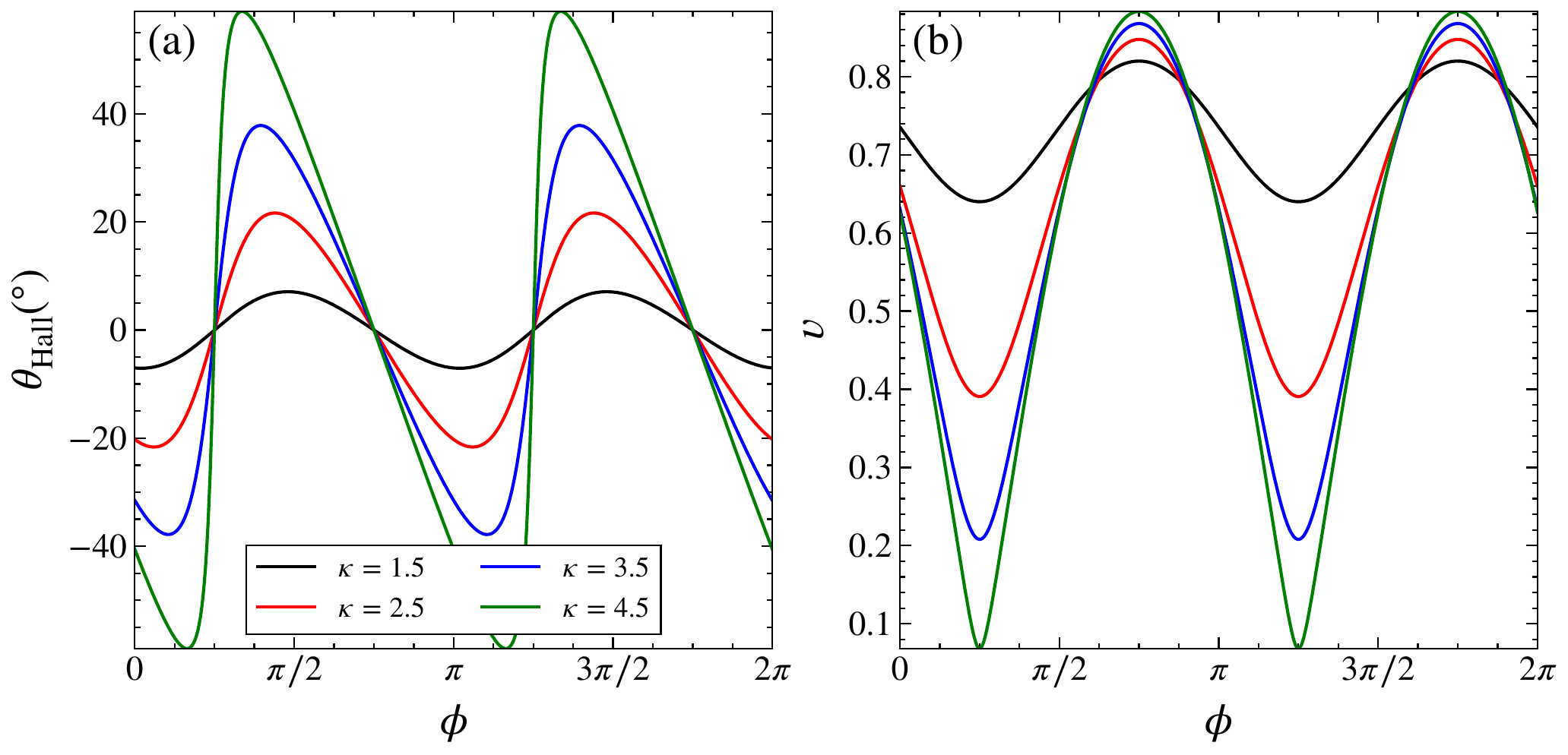}
  \caption{Results of the ATM skyrmion behavior using the
    toy model from Eq.~\ref{eq:4} with parameters
    $G=\mathcal{D}=v_e=1$, $\alpha=0.5$, $\beta=0.01\alpha$
    and different $\kappa=1.5$ (black), 2.5 (red), 3.5 (blue),
    and 4.5 (green).
    (a) ATM skyrmion Hall angle $\theta_\mathrm{Hall}$ vs
    applied current angle $\phi$.
    (b) ATM skyrmion absolute velocity %$v=\sqrt{v_\parallel^2+v_\perp^2}$
    $v$ vs applied current angle $\phi$.
  }
  \label{fig:4}
\end{figure}

In Fig.~\ref{fig:4}(a,b)
we plot the Hall angle $\theta_\mathrm{Hall}$ and the
ATM skyrmion absolute velocity $v$, respectively,
versus $\phi$ obtained using Eqs.~\ref{eq:5}, \ref{eq:6}
with an arbitrary set of parameters.
The Hall angle plot is very similar to the behavior observed in
Figs.~\ref{fig:2}(a) and \ref{fig:3}(a);
however, the velocity curves differ from the velocities
shown in Figs.~\ref{fig:2}(c) and \ref{fig:3}(c).
As previously stated, the goal of
our toy model is to reproduce the asymmetric
behavior present in Figs.~\ref{fig:2}, \ref{fig:3}.
In Fig.~\ref{fig:4}(a),
the current angles $\phi$
for which $\theta_\mathrm{Hall}=0$ match those found in
Figs.~\ref{fig:2}(a) and \ref{fig:3}(a). The sawtooth
asymmetry is also captured by the toy model, especially for
large values of $\kappa$.
Since $\kappa$ dictates how asymmetrical the ATM skyrmion
is, large $\kappa$ values correspond to large
values of $J_2/J_1$.
The velocity curve of the toy model in Fig.~\ref{fig:4}(b)
resembles the curves shown in
Figs.~\ref{fig:2}(c) and \ref{fig:3}(c), with $v\approx 0$ when
$\phi=\pi/4+n\pi$ and peak velocities at $\phi=3\pi/4+n\pi$;
however, the limit of $\kappa \approx 1$ differs from the
limit of $J_2/J_1 \approx 1$.
Comparing these results, we can confidently say that the toy model
correctly captures the
most important behaviors of the ATM skyrmions, namely
(i) the reduced velocity at $\phi=\pi/4+n\pi$,
(ii) the peak velocity at $\phi=3\phi/4+n\pi$,
(iii) the fact that $\theta_\mathrm{Hall}=0$ when $\phi=\pi/4+n\pi/2$,
and (iv)
the sawtooth shape of
$\theta_\mathrm{Hall}$ curve when the asymmetry of the ATM skyrmions is strong.

\section{Effects of Pinning}

\begin{figure}
  \centering
  \includegraphics[width=\columnwidth]{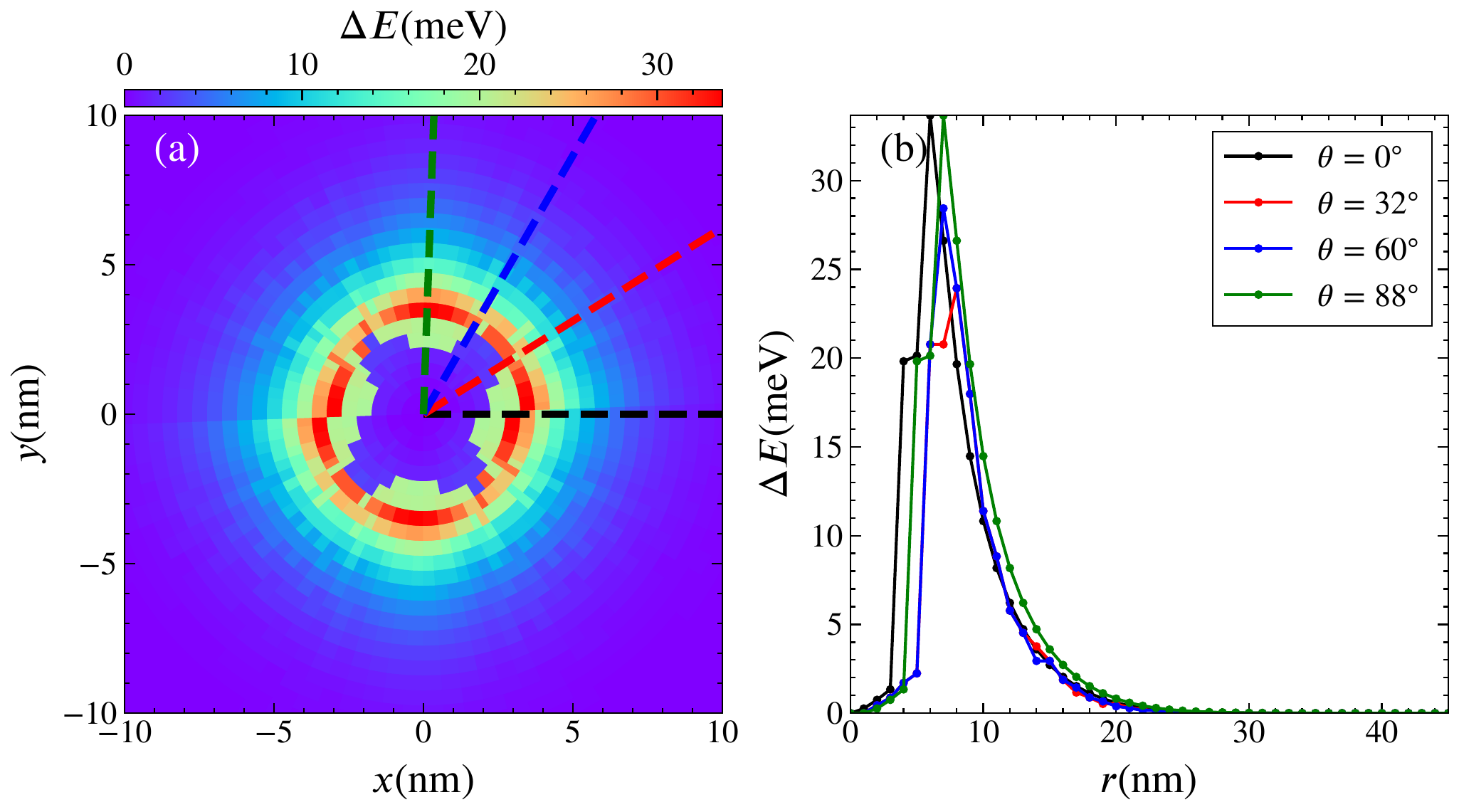}
  \caption{(a) Atomistic simulation calculations of
    the interaction energy of ATM skyrmions
as a function of $y$ vs $x$ position relative to the pinning site center
for circular pinning sites with a radius of $1.5$ nm. The maximum energy occurs just before the skyrmion becomes core pinned. For $r < 15$~nm, the pinning energy is anisotropic due to the anisotropic shape of the skyrmion.
(b) Cross-sections of $\Delta E$ vs $r$ for $\theta = 0.0^\circ$ (black),
$32^\circ$ (red), $60^\circ$ (blue), and $88^\circ$ (green) taken along the dashed lines in panel (a).
}
  \label{fig:5}
\end{figure}

We next consider the effects of pinning on an ATM skyrmion. In Fig.~\ref{fig:5}(a), we show the energy profile as a function of relative position
for an ATM skyrmion interacting with a circular pinning site of radius $1.5$~nm,
calculated using the atomistic model. Figure~\ref{fig:5}(b) shows
the cross section $\Delta E$ of the energy
for slices along $\theta = 0^\circ$, $32^\circ$, $60^\circ$, and $88^\circ$
as a function of the distance $r$ from the pinning site center.
The peak in $\Delta E$ appears at a position just outside the entry of
the skyrmion into the pinning site,
and the dip to
smaller energy for $r < 10$ nm indicates that the skyrmion
core is pinned.
Here, we can see that the anisotropic aspect of the pinning occurs due to the anisotropy of the ATM skyrmion itself;
this effect is stronger for smaller $r$, whereas the behavior becomes
more isotropic for larger $r$.
This anisotropy means that the skyrmions become pinned most easily
at different $r$ values depending upon the relative direction between the
skyrmion core and the pinning site center.
For $r > 15$ nm, the pinning site appears more isotropic to the skyrmion.
Due to the anisotropy of the pinning effectiveness,
the depinning threshold for a trapped skyrmion
differs
for different directions of the applied drive.
In contrast, for ferromagnetic skyrmions,
the pinning energy is isotropic,
and the depinning threshold for different drive directions is
also isotropic.

We next perform atomistic simulations for skyrmions interacting
with a square array of high anisotropy pinning sites.
Previous studies examined ferromagnetic
skyrmions interacting with periodic defect arrays,
and showed that the skyrmion Hall angle
depends on the skyrmion velocity  \cite{Reichhardt15a}.
This effect was
also observed
in simulations and experiments on random pinning arrays
\cite{Reichhardt15, Jiang17, Legrand17, Litzius17, Juge19,Zeissler20} and
arises due to a side-jump effect
in which the Magnus force causes a preferential scattering
off the pinning site.
Slower skyrmions experience a larger jump,
while fast-moving skyrmions are not as affected by the side jump
and move more nearly along the intrinsic Hall angle of
the pinning-free system \cite{Muller15, Reichhardt15a, Fernandes18}.

\begin{figure}
  \centering
  \includegraphics[width=\columnwidth]{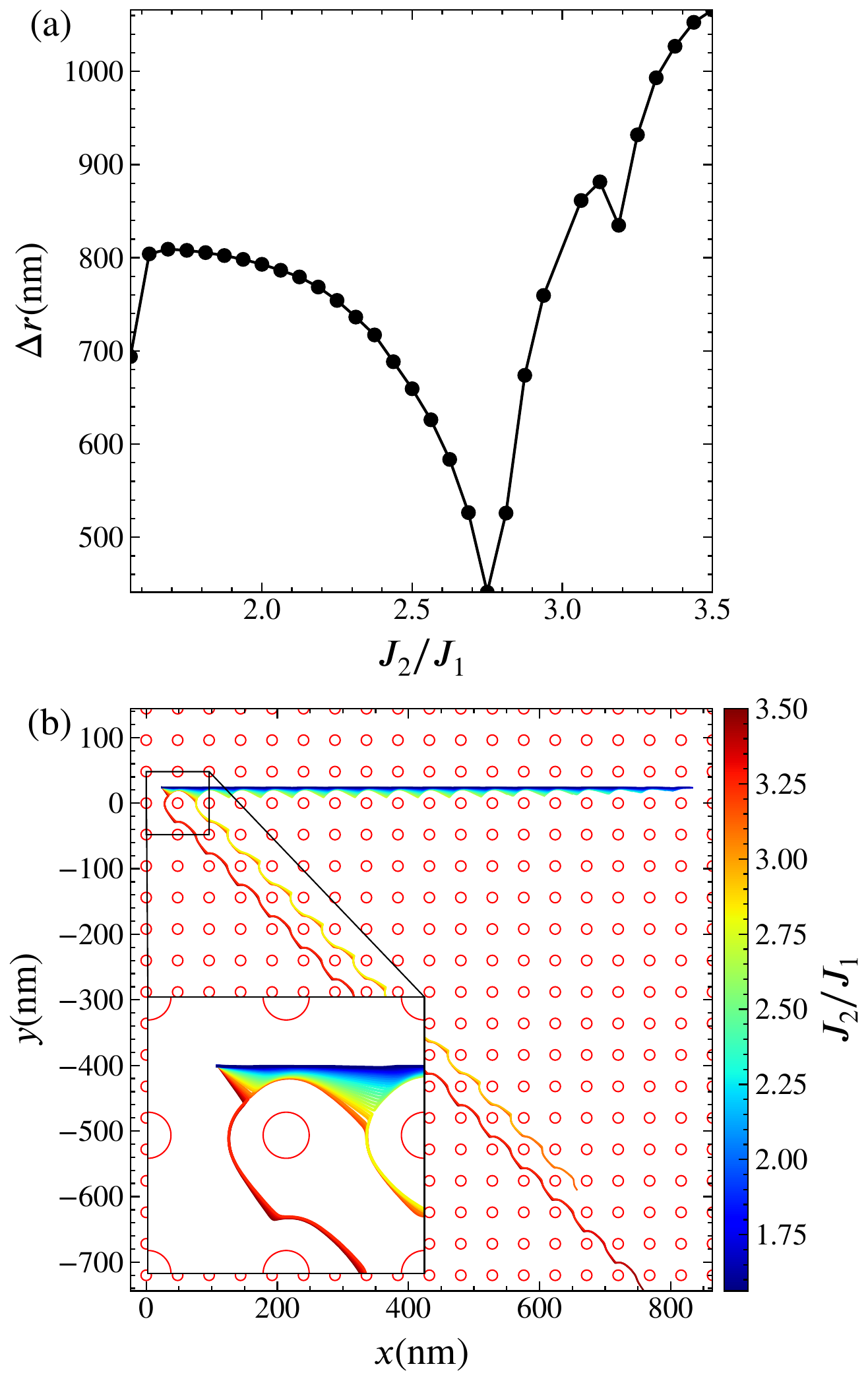}
  \caption{Atomistic simulations of ATM skyrmions interacting with a square
    defect array with different $J_2/J_1$
    at fixed $j=5\times10^{10}$~A~m$^{-2}$ and $\phi=0$.
    (a) Net displacement $\Delta r=\sqrt{\Delta x^2+\Delta y^2}$
    vs $J_2/J_1$.
%    Although the ATM skyrmion velocity
%    increases with $J_2/J_1$, we observe a
%    reduction on the total traveled distance
%    by the ATM skyrmion as a consequence of different
%    $\theta_\mathrm{Hall}$ at different $J_2/J_1$ and
%    interactions with defects.
%    When $J_2/J_1>2.75$, $\Delta r$ starts
%    to increase with $J_2/J_1$, until
%    $J_2/J_1=3.2$ where a second minimum is present.
    (b) ATM skyrmion trajectories as a function of $y$ vs $x$ at different
    $J_2/J_1$ values.
%    The ATM skyrmions flow along three possible
%    trajectories, one with $\theta_\mathrm{Motion}=0$
%    and two with $\theta_\mathrm{Motion}=45^\circ$ as a consequence
%    of interactions with defects.
    The inset shows a detail of the splitting between the three
    observed flow directions.
%    The value of $J_2/J_1$ at which the flow splits
%    matches the value of $J_2/J_1$ at which $\Delta r$
%    has a minimum, associating the traveled distance
%    with changes in the ATM skyrmion flow.
  }
  \label{fig:6}
\end{figure}

In Fig.~\ref{fig:6}(a), we show the
distance traveled by an ATM skyrmion,
$\Delta r=\sqrt{\Delta x^2+\Delta y^2}$,
as a function of $J_2/J_1$
for skyrmions moving over a square defect array at a fixed current
of $j=5\times10^{10}$Am$^{-2}$ and drive direction $\phi=0^{\circ}$.
At lower $J_2/J_1$, the skyrmion motion is locked to $0^{\circ}$, even though there is a finite skyrmion Hall angle for these parameters in the absence
of pinning.
As $J_2/J_1$ increases, the skyrmion velocity in the pin-free case increases,
as shown in Fig.~\ref{fig:3}(c).
When pinning is present, however, the velocity decreases
with increasing $J_2/J_1$, which corresponds
to a decrease in $\Delta r$ in Fig.~\ref{fig:6}(a).
For increasing $J_2/J_1$, the skyrmion Hall angle for the pin-free system also increases, as seen in Fig.~\ref{fig:3}(a),
but when pinning is present,
the skyrmion moves only along the $x$-direction for $J_2/J_1 < 2.7$.
We show this more clearly in Fig.~\ref{fig:6}(b), where we highlight
ATM skyrmion trajectories for different $J_2/J_1$ values.
The motion is locked in the $x$-direction
for $J_2/J_1 \lesssim 2.7$,
while for higher $J_2/J_1$, the skyrmion
travels at a $45^{\circ}$ angle with respect to the pinning array.
This means that just above $J_2/J_1 = 2.7$, the skyrmion
Hall angle is larger than would be expected for the pin-free
system.
This is due to the locking of the skyrmion motion to a substrate
symmetry direction.
It is possible that for less dense pinning or smaller pinning sites, the motion would have locked to a smaller angle, such as one in which the skyrmion
translates by three
substrate lattice constants in $x$ for every
one lattice constant in $y$ to give $\theta = 17.8^{\circ} = \arctan(1/3)$.
The inset in Fig.~\ref{fig:6}(b) shows a detail of
the skyrmion trajectories for increasing $J_2/J_1$.
The skyrmions attempt to move along the $y$-direction, but
strike the next pinning site and are scattered back into motion along
the $x$-direction.
When $J_2/J_1$ and thus the skyrmion Hall angle is large enough,
the skyrmions begin to travel along both the $x$ and $y$ directions because
the skyrmion trajectory is bent strongly enough to pass by the pinning site
without being scattered into the $x$ direction.
%can start moving in both the $x$ and $y$ directions at the
The transition point at which $x$ and $y$ motion first occurs
corresponds to the $J_2/J_1$ value that produces a
minimum in $\Delta r$.
After the transition to motion along $45^{\circ}$,
the skyrmion travels more rapidly with
increasing $J_2/J_1$, as
indicated by the increase
in $\Delta r$ in Fig.~\ref{fig:6}(a) for $J_2/J_1 > 2.7$.
%    When $J_2/J_1>2.75$, $\Delta r$ starts
%    to increase with $J_2/J_1$, until
%    $J_2/J_1=3.2$ where
A dip in $\Delta r$ appears at $J_2/J_1=3.2$, corresponding to the point
at which the initial skyrmion motion passes to the left rather than the right
of the first pinning site in the inset of Fig.~\ref{fig:6}(b).
In previous work on ferromagnetic skyrmions moving over a square periodic array, it was found that the skyrmion motion locks
to symmetry directions such as $0^{\circ}, 45^{\circ}$, and $90^{\circ}$,
as well as other directions corresponding to $\theta = \arctan(n/m)$
with integer $m$ and $n$. Here, the skyrmion
travels by $n$ pinning lattice sites in one direction and
$m$ pinning sites in the other direction,
and the skyrmion velocity
passes through a dip near these directional locking transitions.

We next consider an ATM skyrmion moving over a random array of pinning sites,
which we simulate using our particle model.
We turn to the particle model due to
the computational cost of simulating atomistic ATM skyrmions.
The reduction in the skyrmion velocity by the random pinning array means
that it is necessary to go out to
longer simulation times to observe the behavior,
requiring substantially more computation compared to the periodic pinning
array or the pin-free system.
When considering the square-array we fixed $j$ and $\phi$ while varying
only $J_2/J_1$; however, in order to investigate
the dynamics for random pinning, we must vary both $j$ and $\phi$ while
holding $J_2/J_1$ fixed, increasing the computational cost
even further.
Thus, only simulations with the particle model are practical.
For the particle model,
we consider a two-dimensional system of size $36 \times 36$ with periodic
boundary conditions and $N_p = 100$ repulsive pinning sites with
the Gaussian form
$U(r)=U_0\exp\left(-r^2/a_0^2\right)$,
where $U_0 = 1.0$ and the pin radius $a_0 = 1.0$.
We fix $G=\mathcal{D}=1$, $\alpha=0.5$, $\beta=0.01\alpha$,
and $\kappa=4.5$ in Eq.~\ref{eq:4}.
The driving force ${\bf v}_e$ is ${\bf v}_e=v_e\left(\cos\phi\hat{\bf x}+\sin\phi\hat{\bf y}\right)$.

\begin{figure}
  \centering
  \includegraphics[width=\columnwidth]{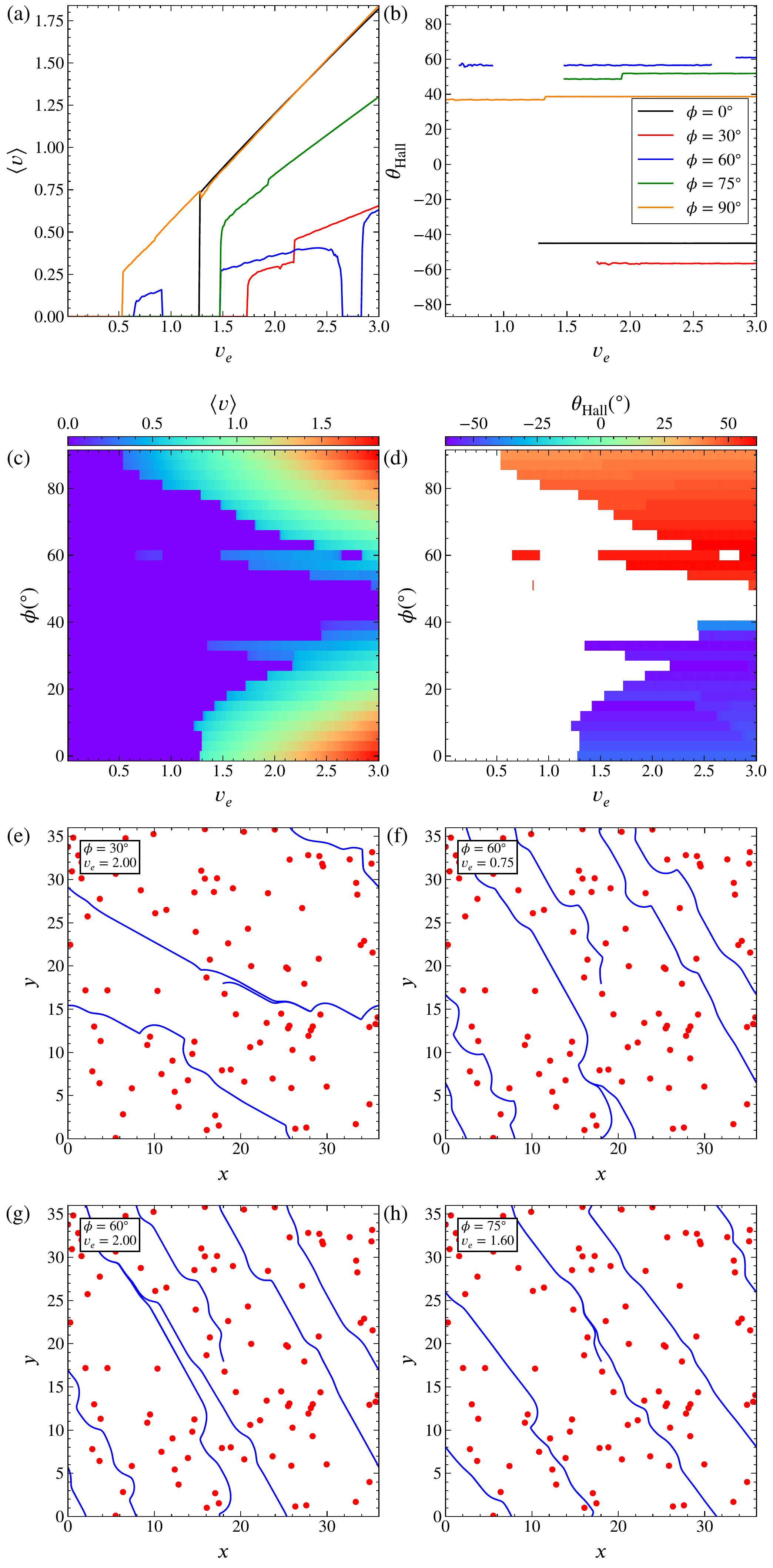}
  \caption{Particle model simulations of ATM skyrmions
interacting with a random array of repulsive pinning sites.
(a) $\langle v \rangle$ vs drive $v_e$ for drive directions $\phi = 0.0^\circ$ (black), $30^\circ$ (red), $60^\circ$ (blue), $75^\circ$ (green), and $90^\circ$ (orange).
(b) The corresponding Hall angle $\theta_\text{Hall}$ vs $v_e$.
(c) Heat map of $\langle v \rangle$ as a function of $\phi$ vs $v_e$. Here, the depinning threshold is anisotropic with a maximum near $\phi = 50^\circ$.
(d) Heat map of $\theta_\text{Hall}$ as a function of $\phi$ vs $v_e$. There is almost no change in $\theta_\text{Hall}$ with varying $v_e$.
(e-h) Skyrmion trajectories (lines) and pinning site locations (red dots) for the ATM skyrmions at (e) $\phi = 30^\circ$ and $v_e = 2.0$, (f) $\phi = 60^\circ$ and $v_e = 7.5$, (g) $\phi = 60^\circ$ and $v_e = 2.0$, and (h) $\phi = 75^\circ$ and $v_e = 1.6$.}
  \label{fig:7}
\end{figure}

\begin{figure}
  \centering
  \includegraphics[width=\columnwidth]{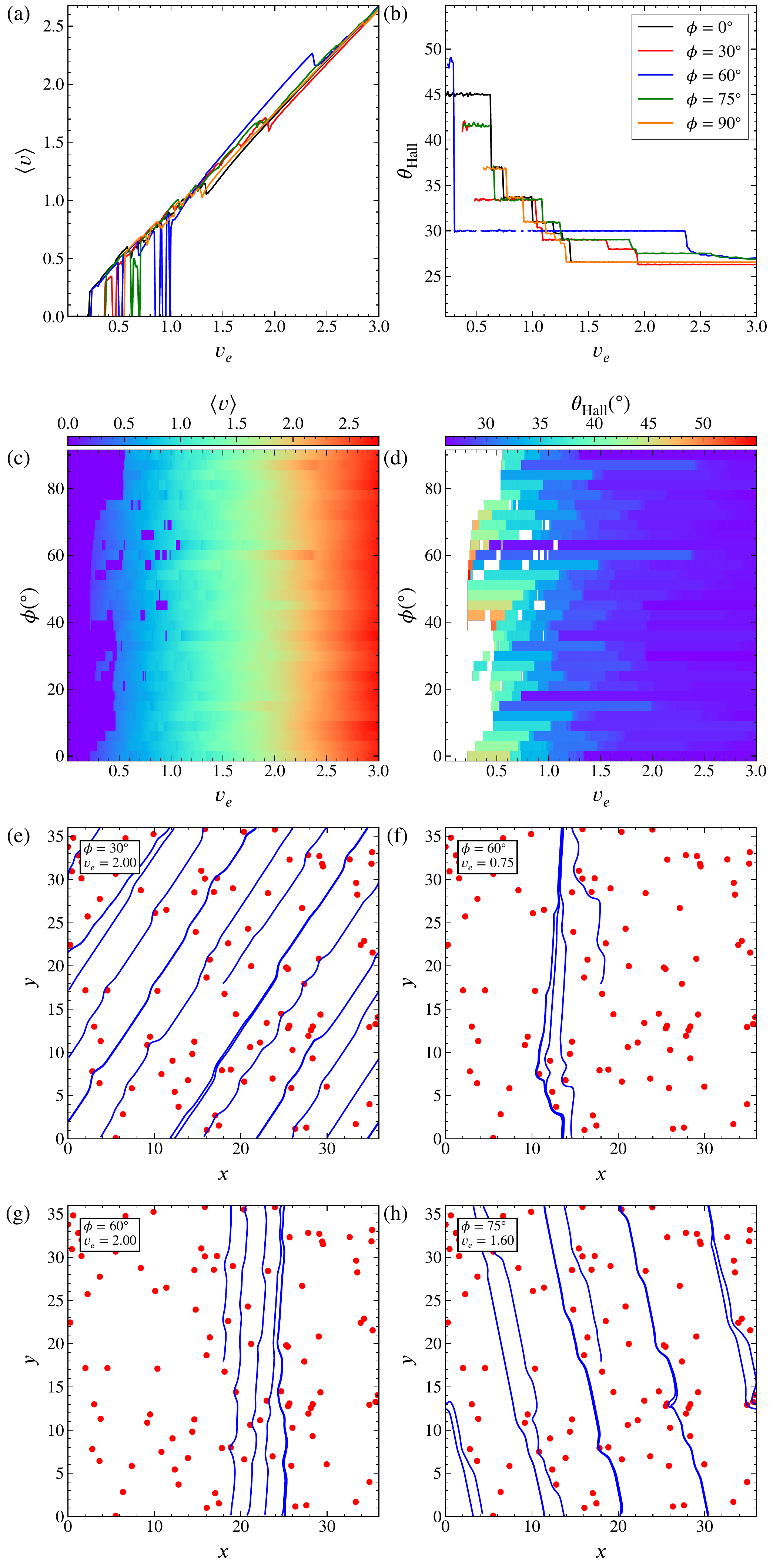}
  \caption{Particle model simulations of FM skyrmions interacting
with a random pinning array of repulsive pinning sites.
(a) $\langle v \rangle$ vs $v_e$ for $\phi = 0.0^\circ$ (black), $30^\circ$ (red), $60^\circ$ (blue), $75^\circ$ (green), and $90^\circ$ (orange).
(b) The corresponding $\theta_\text{Hall}$ vs $v_e$.
(c) Heat map of $\langle v \rangle$
as a function of $\phi$ vs $v_e$. Here, the depinning threshold is isotropic.
(d) Heat map of $\theta_\text{Hall}$ as a function of
$\phi^\circ$ vs $v_e$, showing a drive dependence of $\Theta_\text{Hall}$.
(e-h) Skyrmion trajectories (lines) and pinning site locations (red dots) for the FM skyrmions at (e) $\phi = 30^\circ$ and $v_e = 2.0$,
(f) $\phi = 60^\circ$ and $v_e = 7.5$, (g) $\phi = 60^\circ$ and $v_e = 2.0$,
and (h) $\phi = 75^\circ$ and $v_e = 1.6$.
	}
  \label{fig:8}
\end{figure}

In Fig.~\ref{fig:7}(a), we plot the
velocity $\langle v\rangle$ versus the driving force $v_e$, which is
associated with the velocity of the conduction electrons
\cite{Iwasaki13}, for different current directions $\phi$.
%The different colors represent varied current directions of $0^\circ$ (black),
%$30^\circ$ (red), $60^\circ$ (blue), $75^\circ$ (green), and $90^\circ$ (orange). Here, there are several different features, which include that
Both the depinning threshold and the velocity of the moving skyrmions
change as $\phi$ is varied.
Figure~\ref{fig:7}(b) shows the corresponding skyrmion Hall angle
$\theta_{\rm Hall}$ versus $v_e$ for different $\phi$.
The variation of the skyrmion Hall angle with
$\phi$ in the pin-free system was shown in Fig.~\ref{fig:3}(a)
for the atomistic simulations and
in Fig.~\ref{fig:4}(a) for the particle-based simulation.
In the presence of random pinning, Fig.~\ref{fig:7}(b) indicates that
the AFM skyrmion Hall angle changes very little with increasing $v_{e}$.
For a ferromagnetic skyrmion, the skyrmion Hall angle
is known to have a strong drive dependence when pinning is present,
as has been shown
in continuum \cite{Legrand17}
and particle-based simulations \cite{Reichhardt15, Reichhardt19}
and in experiments
\cite{Jiang17, Litzius17, Zeissler20}.
In Fig.~\ref{fig:8}(a,b), we plot
$\langle v\rangle$ and $\theta_\mathrm{Hall}$ versus
$v_{e}$ from particle simulations of a ferromagnetic (FM) skyrmion
at $\phi = 0^\circ, 30^\circ, 60^\circ, 75^\circ$ and $90^\circ$,
using the same pinning parameters as in Fig.~\ref{fig:7}
for the ATM skyrmion.
We note that these simulations are performed
using the same equation as for the ATM skyrmion, Eq.~\ref{eq:4},
but since the different $G$ skyrmions
do not need to be ``coupled,'' these equations become effectively the same
as those presented
in Ref.~\onlinecite{Iwasaki13}.
In Fig.~\ref{fig:8}(a), the depinning threshold for FM skyrmions is generally lower than that for ATM skyrmions, and the sliding velocity
for $v_{e}$ above the depinning threshold shows far less variation
than in the ATM skyrmion case. In Fig.~\ref{fig:8}(b),
$\theta_\mathrm{Hall}$ decreases with increasing $v_e$,
in agreement with
the expected velocity dependence of
$\theta_\mathrm{Hall}$ found in previous work on non-ATM skyrmions
\cite{Reichhardt15, Legrand17, Jiang17, Litzius17, Zeissler20}.
When $\phi \geq 20^\circ$, $\theta_\mathrm{Hall}$
approaches the intrinsic Hall value at higher $v_e$.

In Fig.~\ref{fig:7}(c), we plot a heat map of the skyrmion
velocity $\langle v\rangle$ for the ATM skyrmions as a function of
$\phi$ versus $v_e$,
and in Fig.~\ref{fig:7}(d) we show the corresponding
heat map of $\theta_\text{Hall}$ as a function of $\phi$ versus $v_e$.
Here, the depinning threshold is anisotropic, with a maximum near $\theta = 45^\circ$ and a minimum at $0^\circ$ and $90^\circ$.
A similar anisotropy appears in the skyrmion velocities.
In contrast, the heat maps of $\langle v\rangle$ and $\theta_\text{Hall}$ as
a function of $\phi$ versus $v_e$ shown in Fig.~\ref{fig:8}(c,d), respectively,
for the FM skyrmions show that
the depinning threshold is isotropic,
with a low depinning threshold that is close to
$v_e = 0.5$ regardless of the drive direction,
and that the velocity response is also isotropic in the moving phase.
For the ATM skyrmions in Fig.~\ref{fig:7}(d), $\theta_\text{Hall}$ is zero in the pinned region and is positive for $\phi > 45^\circ$ and negative for $\phi < 45^\circ$, reflecting the anisotropic nature of the ATM skyrmions.
Additionally, in the moving phase,
the Hall angle of the ATM skyrmions has little dependence on $v_e$
in Fig.~\ref{fig:7}(d), whereas
for FM skyrmions
in Fig.~\ref{fig:8}(d),
$\theta_\text{Hall}$ shows a clear drive dependence with increasing $v_e$,
in agreement with previous work on FM skyrmions \cite{Reichhardt15, Legrand17, Jiang17, Litzius17, Zeissler20}.

In Fig.~\ref{fig:7}(e,f,g,h), we image
the ATM skyrmion trajectories for
$\phi = 30^\circ$ and $v_e = 2.0$, $\phi = 60^\circ$ and $v_e = 0.75$, $\phi = 60^\circ$ and $v_e = 2.0$, $\phi = 75^\circ$ and $v_e = 1.6$, respectively.
For $\phi = 30^\circ$ and $v_e = 2.0$
in Fig.~\ref{fig:7}(e), after an extremely brief transient,
the skyrmion runs across the periodic boundary conditions and follows the
same path during each set of passes across the system.
In Fig.~\ref{fig:7}(f) at $\phi = 60^\circ$ and $v_e = 0.75$,
the motion occurs at a larger angle to the driving direction,
and the skyrmion trajectory has become more random and follows
a different path during each traversal of the system.
For $\phi = 60^\circ$ and $v_e = 2.0$ in Fig.~\ref{fig:7}(g),
the trajectories are very similar to those shown at
$v_e = 0.75$ in Fig.~\ref{fig:7}(f),
reflecting the fact that there is little change in the
Hall angle $\theta_\text{Hall}$ as a function of drive for the ATM skyrmions.
In Fig.~\ref{fig:7}(h) at $\phi = 75^\circ$ and $v_e = 1.6$,
the trajectories look similar to those found
at $\phi = 30^\circ$ and $v_e = 2.0$, but the angle of travel is now larger.

The trajectories of the FM skyrmions appear
in Fig.~\ref{fig:8}(e,f,g,h) at the same values of
$\phi = 30^{\circ}$ and $v_e = 2.0$, $\phi = 60^{\circ}$ and $v_e = 0.75$, $\phi = 60^{\circ}$ and $v_e = 2.0$, and $\phi = 75^{\circ}$ and $v_e = 1.6$, respectively.
For $\phi = 30^{\circ}$ and $v_e = 2.0$,
Fig.~\ref{fig:8}(e) shows that the FM skyrmions travel more rapidly
than the ATM skyrmions, and the motion is more random.
The FM skyrmions also interact with the pinning sites differently,
in that a moving skyrmion  which approaches a pinning site very closely
spirals around the pinning site due to the Magnus force.
This spiraling motion of FM skyrmions around attractive or repulsive pinning sites has been studied in previous computational work
and has been argued to be one of the reasons
that FM skyrmions are weakly pinned compared to particles that
do not have a Magnus force
\cite{Iwasaki13,Nagaosa13,Muller15, Reichhardt15a,Reichhardt15,Fernandes20}.
When the Magnus force is small, the skyrmion moves parallel to the
maximum direction of force,
which points either directly toward or directly away from an
attractive or repulsive pin, respectively.
In Figs.~\ref{fig:7} and ~\ref{fig:8}, the pinning sites are repulsive,
and if the Magnus force is zero, as in the AFM skyrmions of
Fig.~\ref{fig:7}, a skyrmion that moves
toward a pinning site experiences a force in the direction opposite to the
motion and will have difficulty moving around the pinning site.
If there is a cluster of pinning sites
present, the ATM skyrmion can be trapped.
When the Magnus force is finite, as in the FM skyrmions of
Fig.~\ref{fig:8},
there is an additional velocity component perpendicular
to the force from the pinning, permitting
the skyrmion to more easily move around the pinning site,
and permitting even a cluster of pinning sites to be skirted.
As the skyrmion moves around the pinning sites,
it also experiences a side jump effect,
and depending on the sign of the Magnus force,
the side jump preferentially pushes the skyrmion
to one side of the pinning sites,
even when the impact factor would have caused the skyrmion to pass to the
other side of the pinning site in the absence of a
Magnus force.
In Fig.~\ref{fig:8}(e,f,g,h), it is clear that the FM skyrmions
preferentially move around the pinning sites to one side.

\begin{figure}
  \centering
  \includegraphics[width=\columnwidth]{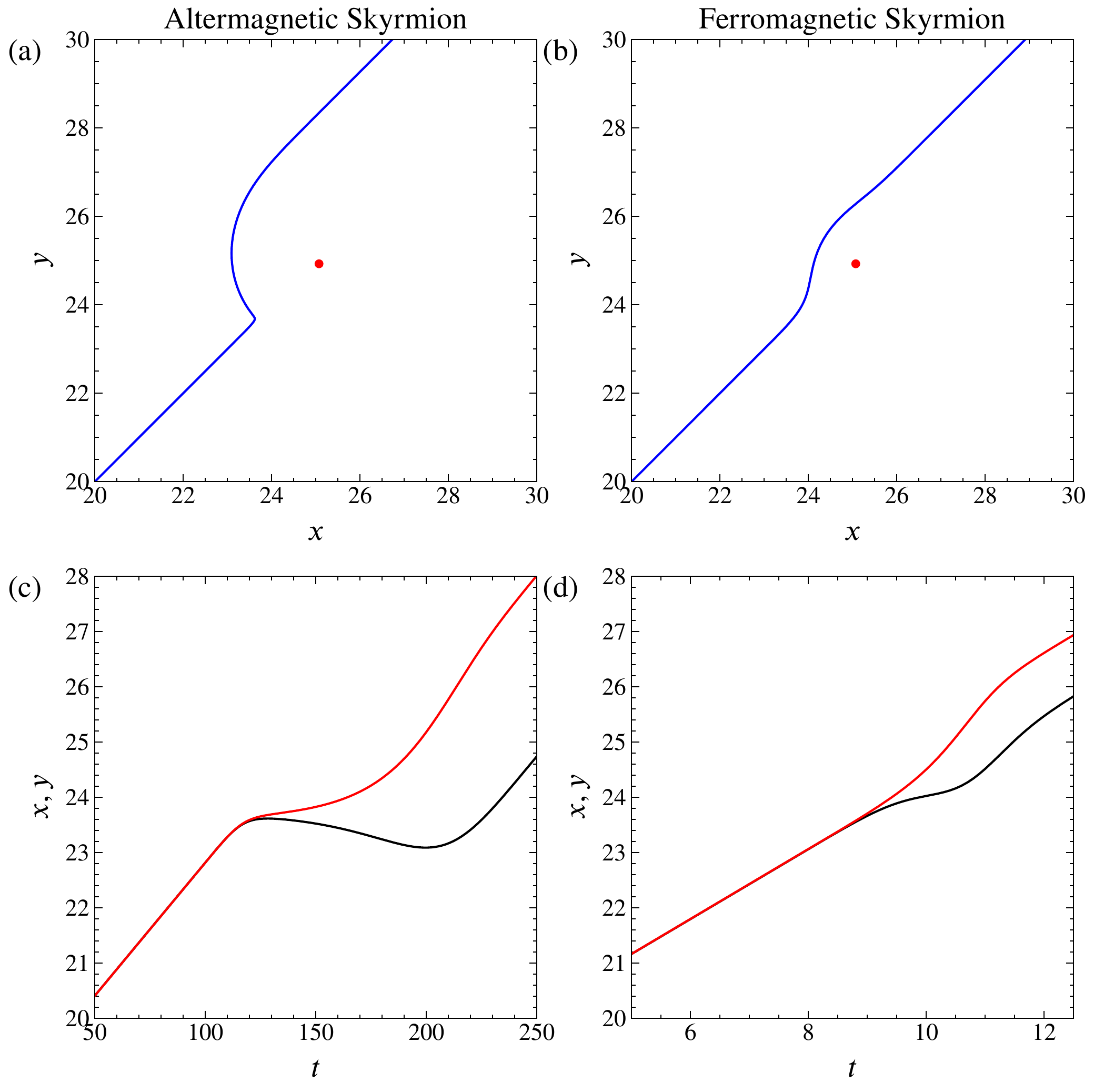}
  \caption{(a) The position and trajectory of an ATM skyrmion interacting with a pinning site, where the skyrmion moves toward the pinning site with an impact parameter of $0.1$.
(b) The position and trajectory of a FM skyrmion under the same conditions. The FM skyrmion does not slow down as much and more easily moves around the pinning site.
(c) The $x$ (red) and $y$ (black) positions of the ATM skyrmion from panel (a) vs time $t$.
(d) The $x$ (red) and $y$ (black) positions of the FM skyrmion from panel (b) vs $t$.
}
  \label{fig:9}
\end{figure}

To show more clearly how the pinning is enhanced for an
ATM skyrmion compared to a FM skyrmion,
in Fig.~\ref{fig:9}(a) we plot the trajectory
of a single ATM skyrmion moving toward a pinning site with an
impact parameter of $0.1$, which is just off-center.
Figure~\ref{fig:9}(c) shows the corresponding $x$ and $y$ position of
the skyrmion as a function of time.
Upon approaching the pinning site,
the ATM skyrmion slows down, and then undergoes
a strong deviation to the side of the pinning site.
Throughout the motion, the minimum distance between the skyrmion
and the pinning site remains relatively large.
Figures~\ref{fig:9}(b) and (d) show the trajectory and time dependent
positions, respectively, of a FM skyrmion under the same conditions,
where we have chosen parameters such that the FM skyrmion approaches
the pinning site at the same angle as the ATM skyrmion.
We find that the FM skyrmion does not slow down as much as it nears the
pinning site, and the minimum distance to
the pinning site is much smaller than in the ATM case.
The FM skyrmion also experiences less deviation as it moves past the
pinning site compared to the ATM skyrmion.
This indicates that the Magnus force reduces the
effectiveness of the pinning for the FM skyrmion compared to the ATM skyrmion.

\section{Discussion}

In this work we considered the motion of individual skyrmions, but
an interesting direction for future exploration
would be to use the atomistic and particle models
to study the collective dynamics of ATM skyrmions.
One open question is how to model the skyrmion-skyrmion interaction potential
for the ATM particle model.
In many-particle models with FM skyrmions,
the skyrmions are considered to experience a short-range,
isotropic repulsion such as a Bessel function
\cite{Lin13,Reichhardt15}
or an exponential repulsion  \cite{Yang22,Ge23}.
In the absence of pinning, the FM skyrmions
form a triangular lattice \cite{Lin13,Reichhardt15},
while when pinning is present, the FM skyrmions can form a
disordered or polycrystalline state  \cite{Gruber26}.
For ATM skyrmions, the skyrmion-skyrmion interactions
should be anisotropic, and the degree of anisotropy would
depend on the different exchange interactions of the two sublattices
as well as on the distance between the skyrmions.

\begin{figure}
  \centering
  \includegraphics[width=\columnwidth]{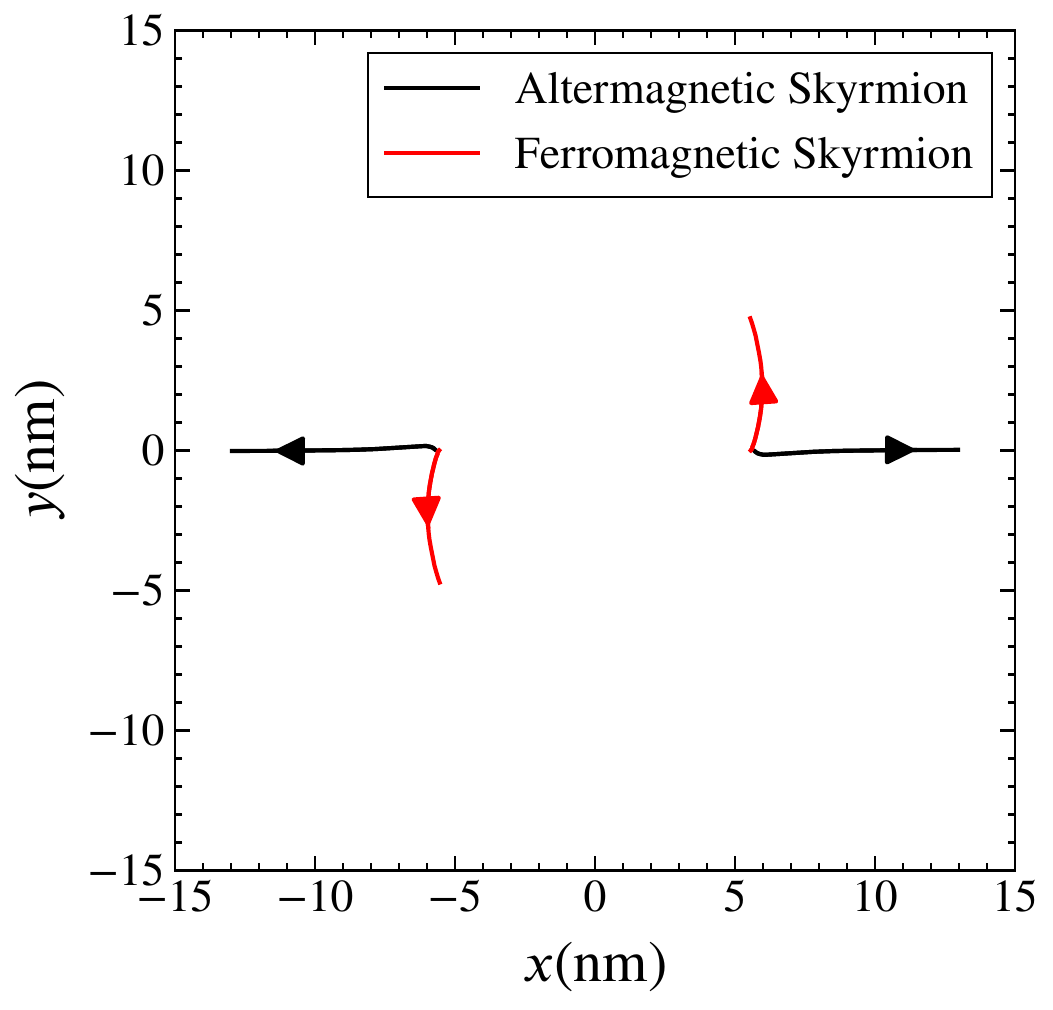}
  \caption{Atomistic simulation of the trajectories of two FM skyrmions (red) and two ATM skyrmions (black) placed near each other.
    The FM skyrmions move outward
    with a spiraling motion due to the Magnus term, and there is skyrmion-skyrmion repulsion. The ATM skyrmions show strong repulsion but the spiraling
    component of the motion is greatly reduced.
  }
  \label{fig:10}
\end{figure}

In Fig.~\ref{fig:10}, we show atomistic simulations of the trajectories
of a pair of FM
skyrmions and, separately, a pair of ATM skyrmions
that are initially placed near each other.
The FM skyrmions move in an outward spiral due to
skyrmion-skyrmion repulsion combined with the Magnus force,
while the ATM skyrmions, though clearly repelled by each other,
move mostly outward due to the weaker Magnus force.
A collective particle model for ATM skyrmions with the same isotropic Bessel function used for FM skyrmions
could be a good approximation for low densities where the anisotropic part of the potential does not overlap.
At higher densities, however, the skyrmion-skyrmion interactions would become more anisotropic,
which could lead to the formation of a square lattice similar to that
found in certain superconducting vortex systems with anisotropic
interactions \cite{Gilardi02,Xie24}.

It is known that under a dc drive, FM skyrmions moving over quenched
disorder can undergo dynamical transitions from a
pinned glass to disordered plastic flow and a
higher-drive triangular lattice,
and that these transitions are independent of the direction of the dc
drive \cite{Reichhardt15}.
For ATM skyrmions, the extent of the
different dynamical phases could depend on the drive direction.
There could be a stronger disordered flow regime for driving
at $45^\circ$ where the depinning threshold is higher,
compared to driving along $0^\circ$ or $90^\circ$.

Another question is the effect of thermal fluctuations on ATM skyrmions. At high enough temperatures,
the system should undergo Brownian-like diffusion, similar to that found for FM skyrmions \cite{Zhao20,Dohi23,Reiger23,Zhang23}.
In the presence of a rough landscape,
the diffusion or thermal creep might be different between the two systems,
however, and the diffusion might be anisotropic for ATM skrymions but
isotropic for FM skyrmions.
Finally, there could be interesting features
in the internal excitation modes of ATM versus FM skyrmions to examine in the
atomistic and micromagnetic models.
The low skyrmion Hall angle for ATM skyrmions is appealing for
possible future applications,
but the fact that ATM skyrmions are more susceptible to pinning
compared to FM skyrmions could limit their usefulness.

\section{Summary}

%CIJOL WORKING
Using atomistic simulations, we have studied the driven dynamics of skyrmions in an altermagnet (ATM). We find that the ATM skyrmion velocity and Hall angle are anisotropic, and that as the ratio of the exchange constants $J_{2}/J_1$ increases, the skyrmion becomes
increasingly ellipsoidal in shape and shows increasing anisotropy in the Hall angle and velocity. We also propose a particle model for the AFM skyrmion that is simpler than the one presented by Jin {\it et al.} \cite{Jin24}. In our model, we consider the dissipation components to vary with the sub-lattice directions. Despite the simplicity of the model, it captures the same general features as the atomistic model, including the variation in the skyrmion Hall angle and velocity for changing drive direction.
Using an atomistic model, we find that when an ATM skyrmion interacts with a pinning site, the pinning is anisotropic at shorter distances due to the shape of the skyrmion. For an ATM skyrmion interacting with a periodic array of pinning sites for a fixed drive direction, we find that as the ratio $J_{2}/J_1$ increases, the skyrmion motion transitions between symmetry directions
from being locked along $0^\circ$ to
being locked along $45^{\circ}$, with a dip in velocity
occurring at the transition. For an ATM skyrmion moving over random pinning, we find that the depinning threshold and velocity of the moving skyrmion are strongly anisotropic, and the skyrmion Hall angle shows little change with increasing drive. In contrast, for FM skyrmions, the depinning threshold is isotropic, and the skyrmion Hall angle has a strong drive dependence due to the presence of the Magnus force, which creates a side jump effect.
In general, ATM skyrmions are more strongly affected by pinning than FM skyrmions due to the reduction of the Magnus force. We also discuss several future directions, such as collective and thermal effects for ATM skyrmions.

\acknowledgments

This work was supported by the US Department of Energy through the Los Alamos
National Laboratory. Los Alamos National Laboratory is operated by
Triad National Security, LLC, for the National Nuclear Security
Administration of the U. S. Department of Energy (Contract
No. 892333218NCA000001).
%
%J.C.B.S acknowledges partial funding from Fundação de Amparo à
%Pesquisa do Estado de São Paulo - FAPESP, grants 2023/17545-1
%2022/14053-8.
%
We would like to thank Dr. Pablo Venegas for providing the computational
resources used in this work via FAPESP (Grant: 2024/02941-1).

\bibliography{mybib}
\end{document}